\documentclass[fleqn,12pt,twoside]{article}
\usepackage{espcrc1}

\readRCS
$Id: espcrc1.tex,v 1.2 2004/02/24 11:22:11 spepping Exp $
\usepackage{graphicx}
\usepackage{amsmath}
\usepackage{amsfonts,amsbsy}

\newcommand{\slvarepsilon}{\raise.15ex\hbox{$/$}\kern-.53em\hbox{$\varepsilon$}}
\newcommand{\slL}{\raise.15ex\hbox{$/$}\kern-.53em\hbox{$L$}}
\newcommand{\slP}{\raise.15ex\hbox{$/$}\kern-.53em\hbox{$P$}}
\newcommand{\slp}{\raise.1ex\hbox{$/$}\kern-.63em\hbox{$p$}}
\newcommand{\slq}{\raise.1ex\hbox{$/$}\kern-.53em\hbox{$q$}}
\newcommand{\slv}{\raise.1ex\hbox{$/$}\kern-.63em\hbox{$v$}}
\newcommand{\slR}{\raise.15ex\hbox{$/$}\kern-.53em\hbox{$R$}}
\newcommand{\slQ}{\raise.15ex\hbox{$/$}\kern-.53em\hbox{$Q$}}
\newcommand{\slK}{\raise.15ex\hbox{$/$}\kern-.53em\hbox{$K$}}
\newcommand{\slk}{\raise.15ex\hbox{$/$}\kern-.53em\hbox{$k$}}
\newcommand{\slSigma}{\raise.15ex\hbox{$/$}\kern-.53em\hbox{$\Sigma$}}
\newcommand{\slcalP}{\raise.15ex\hbox{$/$}\kern-.63em\hbox{$\cal P$}}
\newcommand{\slA}{\raise.15ex\hbox{$/$}\kern-.73em\hbox{$A$}}
\newcommand{\slbfA}{\raise.15ex\hbox{$/$}\kern-.73em\hbox{${\imb A}$}}
\newcommand{\slpartial}{\raise.15ex\hbox{$/$}\kern-.53em\hbox{$\partial$}}
\newcommand{\sla}{\raise.15ex\hbox{$/$}\kern-.53em\hbox{$a$}}
\newcommand{\slb}{\raise.15ex\hbox{$/$}\kern-.53em\hbox{$b$}}
\newcommand{\slc}{\raise.15ex\hbox{$/$}\kern-.53em\hbox{$c$}}
\newcommand{\slC}{\raise.15ex\hbox{$/$}\kern-.63em\hbox{$C$}}

\def\p{{\boldsymbol p}}
\def\q{{\boldsymbol q}}
\def\l{{\boldsymbol l}}
\def\k{{\boldsymbol k}}

\def\x{{\boldsymbol x}}

\def\y{{\boldsymbol y}}

\def\0{{\boldsymbol 0}}

\def\wt{\widetilde}

\title{Violation of $k_\perp$ factorization in quark production from the
Color Glass Condensate}

\author{%
H.~Fujii,%
\address{Institute of Physics, University of Tokyo, Komaba, Tokyo 153-8902, Japan} \,
F.~Gelis%
\address{CEA/DSM/SPhT, 91191 Gif-sur-Yvette cedex, France} 
and
R.~Venugopalan%
\address{Physics Department, Brookhaven National Laboratory, Upton, New York 11973, USA}
}
       
\begin{document}

\maketitle

\begin{abstract}
We examine the violation of the $k_\perp$ factorization
approximation for quark production in high energy proton-nucleus collisions.
We comment on its implications for the open charm and quarkonium production in 
collider experiments.
\end{abstract}

\section{Introduction}

Semi-hard processes, where $\sqrt{s} \gg m_{q_\perp} \gg \Lambda_{\rm QCD}$,
contribute significantly to  particle production
in high-energy collider experiments due to the large density
of the small-$x$ gluons.
The $k_\perp$ factorization formalism\cite{ktfact} systematically 
resums corrections of $(\alpha_s \ln(s/q_\perp^2))^n$ from 
gluon branchings in perturbative QCD.
In this framework, the particle production cross-section
is expressed as a convolution of a hard matrix element and {\em unintegrated}
distributions of gluons in the hadrons with definite transverse momentum
$\k_{i \perp}$ and longitudinal fraction $x_i$ in each projectile hadron
($i$=1, 2).

Multiple-scattering (higher twist) effects  become important at small $x$ due to 
the large density of small-$x$ gluons. It is expected to 
be the origin of the Cronin enhancement
and $p_\perp$ broadening of hadrons observed in nuclear experiments. 
It is also relevant for the nuclear suppression of quarkonium production.

The simplest situation for studying the impact of higher twist effects 
on $k_\perp$ factorization is in proton-nucleus (pA) collisions, 
wherein the proton is dilute and the nucleus is dense.
The $k_\perp$ factorization formalism was examined in the color glass condensate
framework\cite{CGC}.  It is shown that 
the factorization is recovered when one keeps only the terms that
are of the lowest order in the charge sources $\rho_{p,A}$ of the projectiles\cite{GelisV1}.
The cross-sections at the leading order in $\rho_p$, but at all orders 
in the dense source $\rho_{A}$ of the nucleus are obtained analytically. 
Gluon production by the ``2-to-1'' processes is shown to be  
$k_\perp$-factorizable \cite{gluons1,BlaizGV1,gluons2}
whereas the quark production is generally not\cite{BlaizGV2,Tuchin,NikolS1}.

Here we report the numerical estimates for the $k_\perp$ factorization
breaking in quark production within the McLerran-Venugopalan (MV) model\cite{FGV1}.
We briefly discuss  open charm production and  quarkonium suppression
in  pA collisions in this framework.

\section{Violation of $k_\perp$ factorization in quark pair production}

The quark pair production cross-section is obtained as\cite{BlaizGV2}:
\begin{eqnarray}
\frac{d\sigma}{d^2\p_\perp d^2\q_\perp dy_p dy_q} &\!\!\!\!=&\!\!\!\!
\frac{\alpha_s^2 N}{8\pi^4 (N^2-1)}
\int\limits_{\k_{1\perp},\k_{2\perp}}
\frac{\delta(\p_\perp+\q_\perp-\k_{1\perp}-\k_{2\perp})}
{\k_{1\perp}^2 \k_{2\perp}^2}
\nonumber\\
&&\!\!\!\!\!
\times\Bigg\{
\int\limits_{\k_\perp,\k_\perp^\prime}
\!\!\!\!\!
{\rm tr}_{\rm d}
\Big[(\slq\!+\!m)T_{q\bar{q}}(\slp\!-\!m)
\gamma^0 T_{q\bar{q}}^{\prime\dagger}\gamma^0\Big]
\phi_{_A}^{q\bar{q},q\bar{q}}
(\k_{2\perp}; \k_\perp, \k_\perp^\prime)
\nonumber\\
&&\;\;
+\int\limits_{\k_\perp}
\!
{\rm tr}_{\rm d}
\Big[(\slq\!+\!m)T_{q\bar{q}}(\slp\!-\!m)
\gamma^0 T_{g}^{\dagger}\gamma^0 + {\rm h.c.}\Big]
\phi_{_A}^{q\bar{q},g}
(\k_{2\perp};\k_\perp)
\nonumber\\
&&\qquad
+{\rm tr}_{\rm d}
\Big[(\slq\!+\!m)T_{g}(\slp\!-\!m)\gamma^0 T_{g}^{\dagger}\gamma^0\Big]
\phi_{_A}^{g,g}(\k_{2\perp})
\Bigg\}
\varphi_p(\k_{1\perp})\; ,
\label{eq:cross-section}
\end{eqnarray}
where the explicit forms for the Dirac matrices
$T_{q\bar{q}}(\k_{1\perp},\k_{\perp})$ and
$T_{g}(\k_{1\perp})$ are given in \cite{BlaizGV2}.
Here 
$
\varphi_p(\l_\perp)\equiv 
({\pi^2 R_p^2 g^2 / l_\perp^2})
\; {\rm F.T.}\, 
\left<\rho_p^a(\0)\rho_p^a(\x_\perp)\right>\; 
$
is the unintegrated gluon distribution for the proton,
and F.T.\ denotes the Fourier transformation.  
One needs, however, {\em three} nuclear distributions defined as
(see Eqs.~(42), (43) and (45) in \cite{BlaizGV2})
\begin{eqnarray}
&&\phi_{_{A}}^{g,g}(\l_{\perp})\equiv \frac{\pi^2 R_{_A}^2 l_\perp^2}{g^2 N}
\; {\rm F.T.}\; 
\,{\rm tr}\,\left<U(\0)U^\dagger(\x_\perp)\right>\; ,
\nonumber\\
&&
\phi_{_A}^{q\bar{q},g}(\l_\perp;\k_\perp)\equiv
\frac{2\pi^2 R_{_A}^2 l_\perp^2}{g^2 N}
\; {\rm F.T.}\; 
{\rm tr}
\left<
{\wt U}(\x_\perp)t^a {\wt U}^\dagger(\y_\perp) t^b U_{ba}(\0)
\right> \; , 
\nonumber\\
&&\phi_{_A}^{q\bar{q};q\bar{q}}(\l_\perp;\k_\perp;\k_\perp^\prime)\equiv 
\frac{2\pi^2 R_{_A}^2 l_\perp^2}{g^2 N}
\; {\rm F.T.}\; 
{\rm tr}
\left<{\wt U}(\0)t^a {\wt U}^\dagger(\y_\perp){\wt U}(\x_\perp^\prime)t^a {\wt U}^\dagger(\y_\perp^\prime)
\right> \, ,
\label{eq:corr-nucleus}
\end{eqnarray}
where $U$ and $\wt U$ denote the path-ordered exponentials of the
gauge fields in the nucleus in the adjoint and fundamental representations, respectively,
and describe the multiple scatterings of the gluon and the quarks.
The average $\left< \cdots \right>$ is taken
over the Gaussian distribution of the color charge sources characterized
by the saturation scale $Q_s^2$.

$k_\perp$ factorization is violated by the transverse structure
of the quark pair probed by the momentum ${\k_\perp}^{(\prime)}$ from the nucleus since each 
quark from the pair can resolve and interact with several gluons from the nucleus.
If any of the transverse masses $m_{q_\perp}$ and $m_{p_\perp}$ of the produced quarks
is large compared with the typical rescattering scale, $Q_s$,
we can neglect $\k_\perp^{(\prime)}$ in $T_{q\bar{q}}(\k_{1\perp},\k_{\perp}^{(\prime)})$ 
and recover the $k_\perp$ factorized formula thanks to the sum rule for $\phi_A$'s;
$
\int_{\k_\perp,\k_\perp^\prime}\phi^{q\bar q,q\bar q}_A
= 
\int_{\k_\perp}\phi^{q\bar q,g}_A
=
\phi^{g,g}_A
$\; .

\begin{figure}[tb]
\begin{minipage}[t]{75mm}
\centerline{\includegraphics[width=7cm]{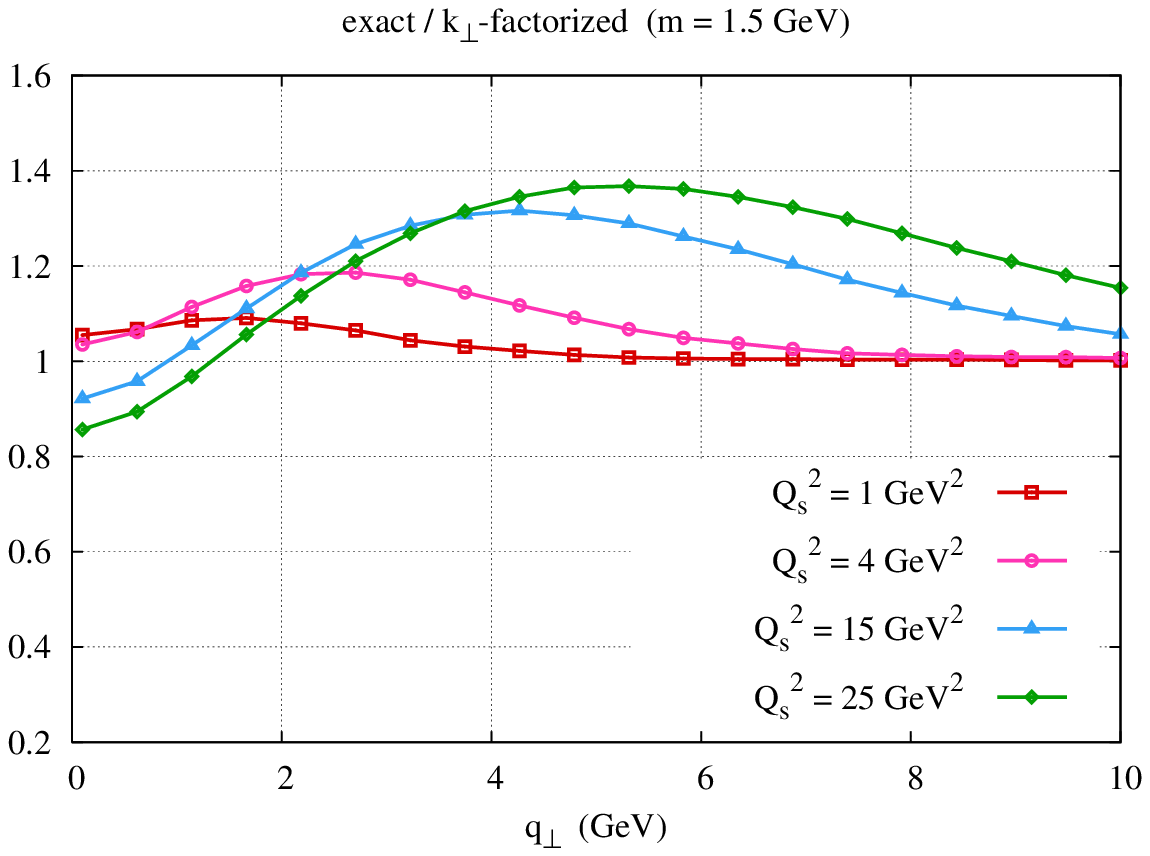}}
\caption{Breaking of  $k_\perp$ factorization in  single charm quark production.}
\label{fig:ktbreakC}
\end{minipage}
\hspace{\fill}
\begin{minipage}[t]{75mm}
\centerline{\includegraphics[width=7cm]{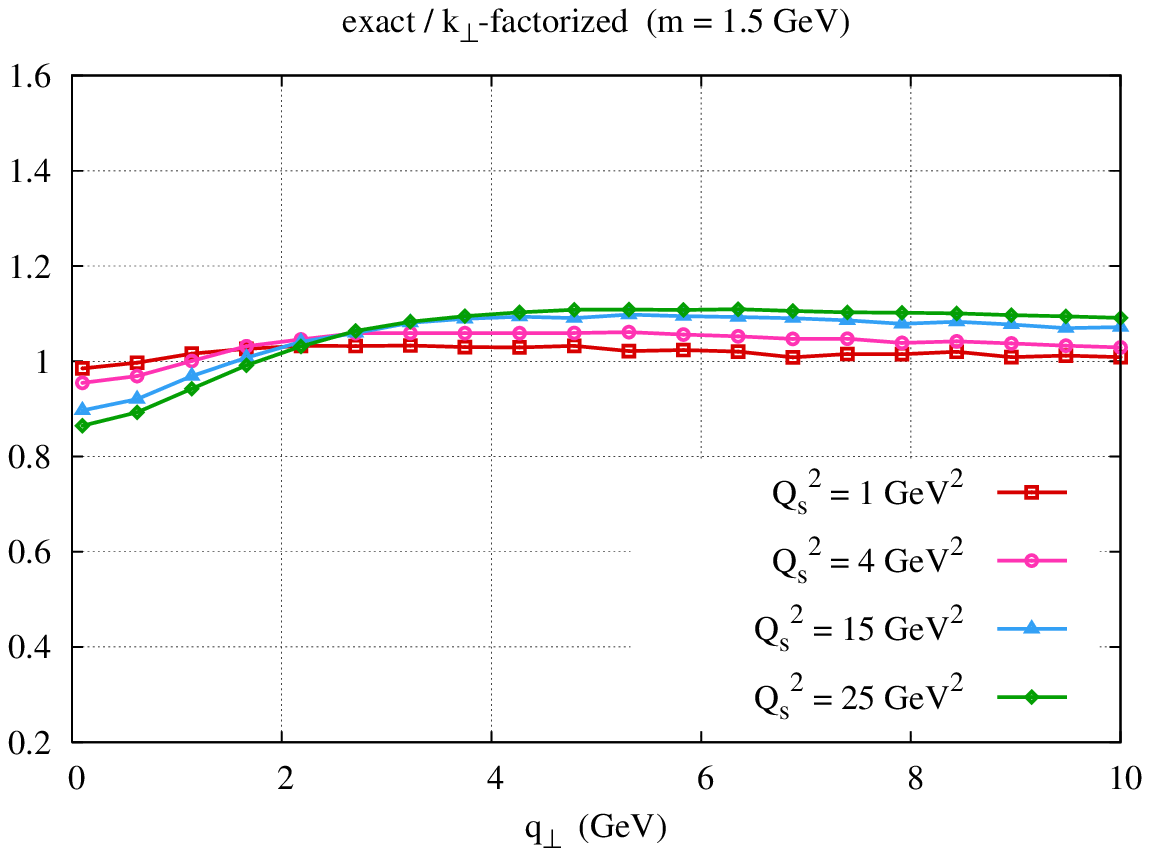}}
\caption{The same as in Fig.~1 but in the nonlocal Gauss model.}
\label{fig:ktbreakCnonG}
\end{minipage}
\begin{minipage}[t]{75mm}
\centerline{\includegraphics[width=7cm]{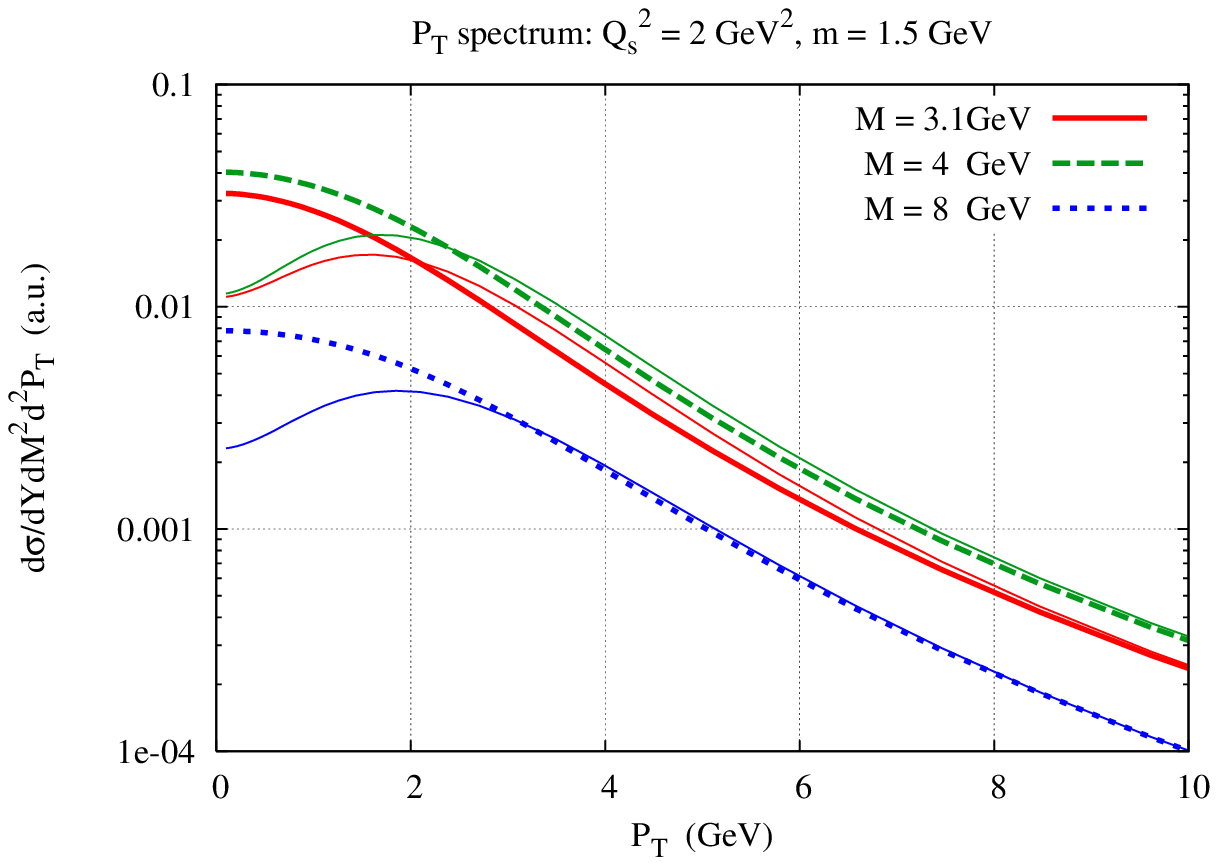}}
\caption{$P_\perp$ spectrum of the quark pair with fixed invariant mass $M$.}
\label{fig:Ptspectrum}
\end{minipage}
\hspace{\fill}
\begin{minipage}[t]{75mm}
\centerline{\includegraphics[width=7cm]{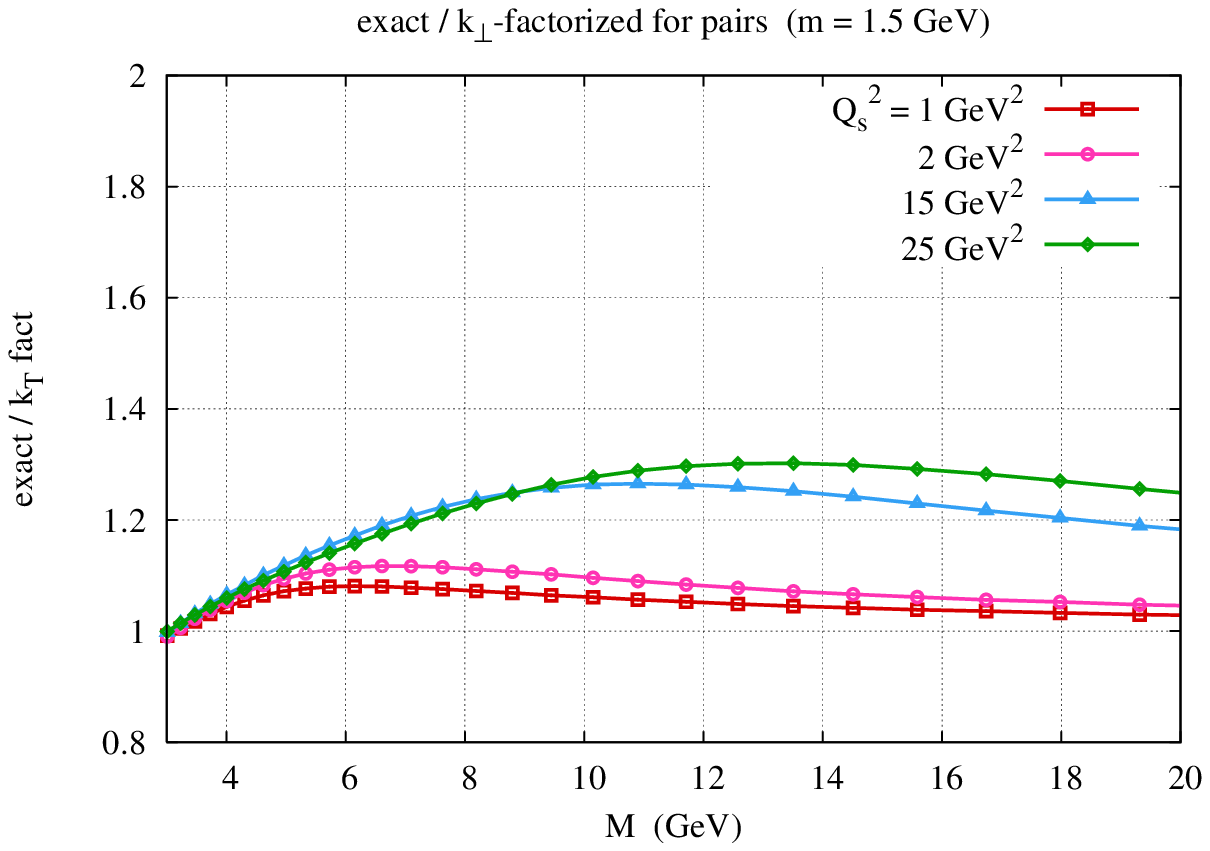}}
\caption{Breaking of  $k_\perp$ factorization in  charm quark pair production.}
\label{fig:ktbreakCpair}
\end{minipage}
\end{figure}

In Fig.~1 we compare the exact result with the $k_\perp$ factorized approximation for  single 
charm quark production. 
The breaking is relatively small for the saturation momentum $Q_s^2$=1 GeV$^2$, which may be
the relevant scale for RHIC at central rapidity. 
At $Q_s^2$=15, 25 GeV$^2$ (corresponding to very forward rapidities in the proton fragmentation 
region at RHIC and LHC)  the correction can be as large as 40\% at 
$q_\perp\sim Q_s$. For the bottom quark production the violation is smaller.
To assess the model-dependence of our results, we compute them now, shown in Fig.~2,  with 
a non-local Gaussian model  known to be the asymptotic solution of renormalization equations for 
$x$ evolution\cite{IancuIM2};  non-linear evolution effects reduce the magnitude of the violation 
of $k_\perp$ factorization.

In Fig.~3 shown is the total $P_\perp$ distribution of the charm quark pair with
the fixed invariant masses $M$=3.1, 4, 8 GeV. 
In the $k_\perp$ factorized approximation (thin curves),
either quark or antiquark exchanges all the momentum from
the nucleus and we see the bump structure near $Q_s$, reflecting
the gluon distribution of the nucleus.
The bump is smeared out due to multiple scatterings of both the quark and antiquark
in the full formula.
Integrating over $P_\perp$, we show in Fig.~4, the magnitude of factorization breaking 
in the invariant mass spectrum of the pair.

\section{Phenomenology}

We study the importance of small-$x$ distributions in D meson production
by convoluting the single quark spectrum with an appropriate fragmentation function\cite{KharzT1}. 
We find, however, the production spectrum is determined 
not by the quark distribution with 
$q_\perp {\raise-.55ex\hbox{{$\sim$}\kern-.8em\raise.80ex\hbox{$<$}}} Q_s$,
but largely by the tail part $\propto 1/q_\perp^4$ of the MV model. 
Moreover, in order to assess the rapidity dependence of open
charm production, the $x$-dependence of the unintegrated gluon
distributions should be taken into account, which requires going
beyond the MV model. Our results on open charm production will be
reported elsewhere\cite{FGV2}.

The $Q_s^2$-dependence of the pair spectrum (divided by the charge density
$\mu_A^2$) is diplayed in Fig.~5. 
At larger $M$, where the high-density effects are diminished,
all curves converge to a single one.
The multiple scatterings of the pair quarks suppress the yield in the low $M$ region. (The 
overall cross-section is of course enhanced with increasing $Q_s^2$.)
One can get an idea about the normal suppression of the quarkonium production in
the pA collisions, relying on the color evaporation picture.
We show the nuclear modification ratio,
$R_{pA}$, for the pairs with $M$ less than the open charm threshold $2M_D$,
as a function of $Q_s^2$.
The suppression pattern fits the form $1/(Q_s^2)^\alpha$ with $\alpha \sim 0.42$,
and not the frequently assumed exponential form. 
One should note here that $Q_s^2 \sim A^{1/3}$ in the MV model.

\begin{figure}
\begin{minipage}[t]{75mm}
\centerline{\includegraphics[width=6.5cm]{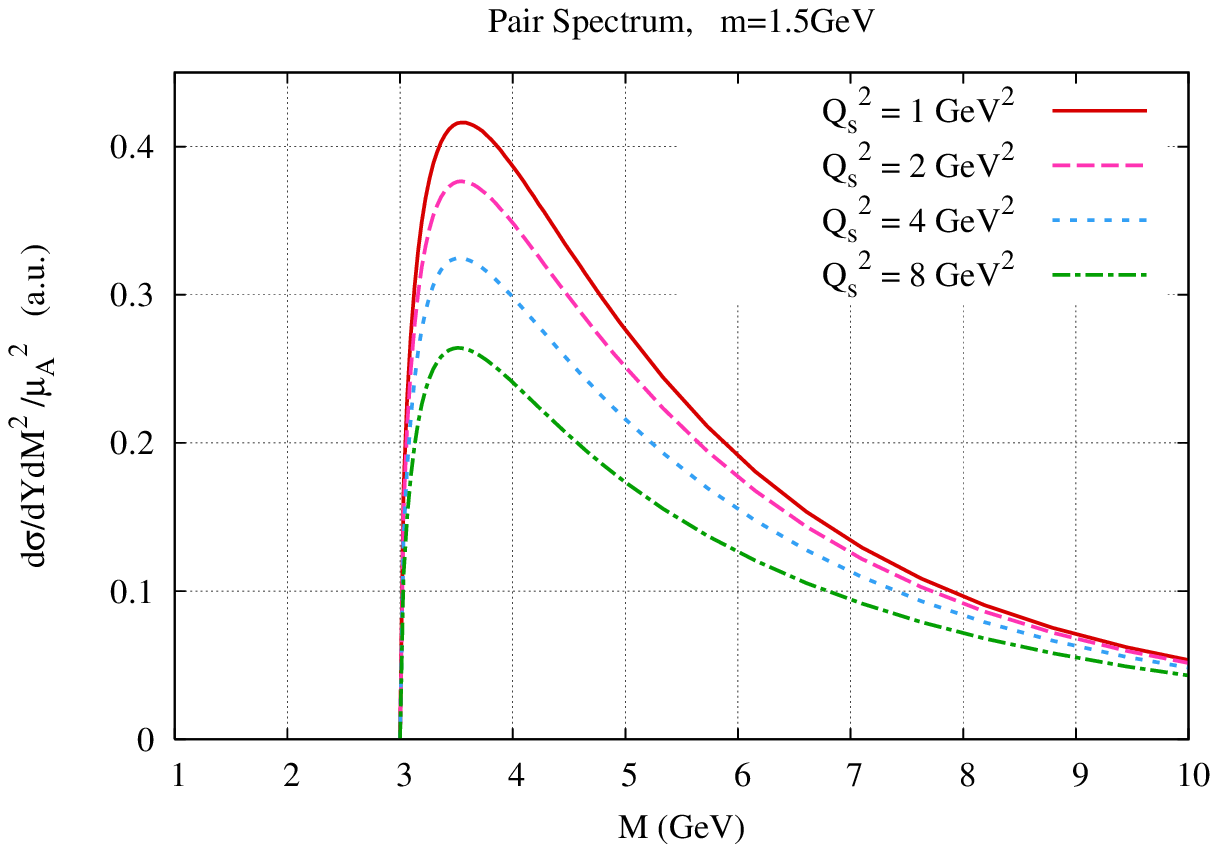}}
\caption{$Q_s^2$ dependence of the charm pair production.}
\label{fig:pairvsqs2}
\end{minipage}
\hspace{\fill}
\begin{minipage}[t]{75mm}
\centerline{\includegraphics[width=6.5cm]{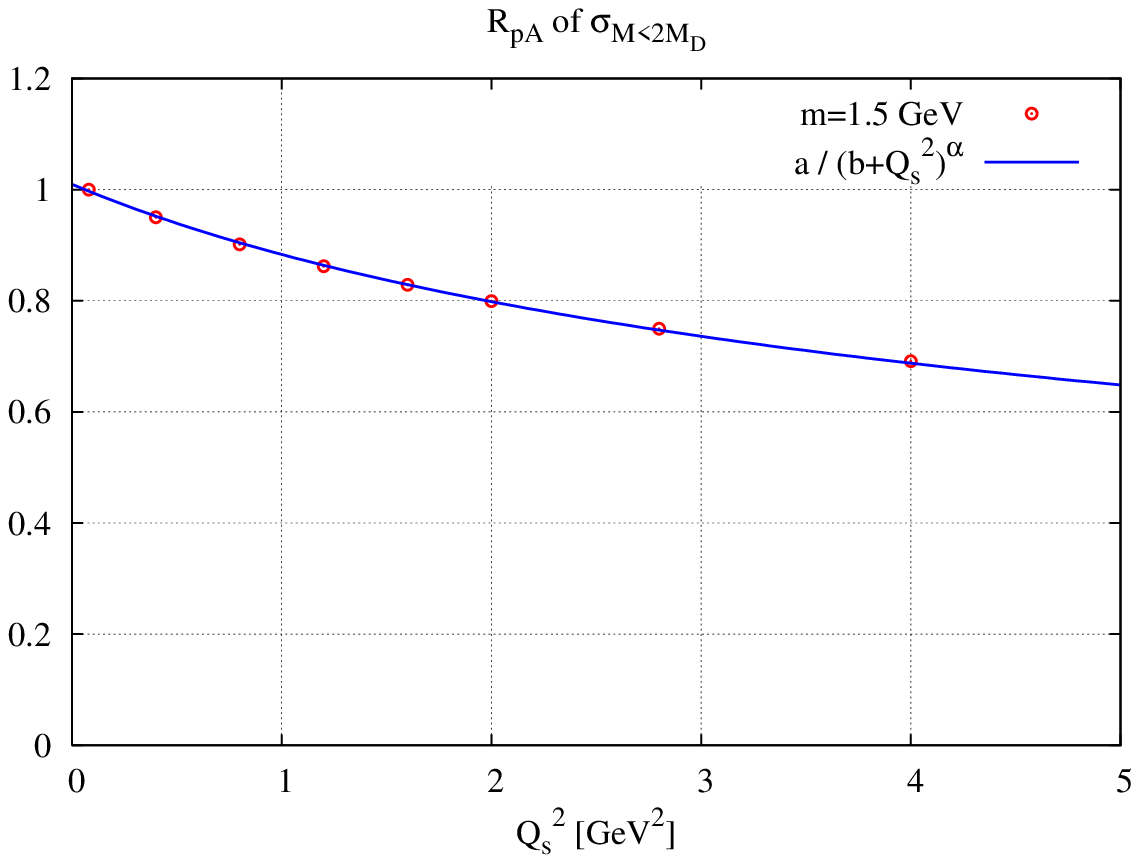}}
\caption{Suppression of low mass pairs in pA collisions.}
\label{fig:onium}
\end{minipage}
\end{figure}

\end{document}